\documentclass[12pt]{article}
\usepackage{times}
\usepackage{geometry}
\usepackage[utf8]{inputenc}
\usepackage{enumitem,amssymb}
\usepackage{ragged2e}
\usepackage{pifont}
\usepackage{outlines}
\usepackage{xcolor}
\usepackage{wrapfig}
\usepackage{graphicx}
\usepackage{siunitx}

\usepackage[]{natbib} 

\usepackage[citecolor=blue, colorlinks=True]{hyperref}
\usepackage{multicol}

\begin{document}
{\raggedright
\LARGE
\textbf{Hubble Science in the 2030s White Paper:}\\ High-Contrast Optical and UV Spectroscopy with HST/STIS \linebreak
\normalsize

\textbf{Principal Author:}

Name:	Kimberly Ward-Duong
 \linebreak						
Institution:  Smith College
 \linebreak
Email: kwardduong@smith.edu
 \linebreak
 
\textbf{Co-authors:} \\ 
John Debes (Space Telescope Science Institute)\\
Jonathan Aguilar (Space Telescope Science Institute)\\
Thayne Currie (University of Texas - San Antonio)\\
Jamie Lomax (United States Naval Academy)\footnote{The views expressed in this white paper are those of the authors and do not reflect the official policy or position of the U.S. Naval Academy, Department of the Navy, the Department of Defense, or the U.S. Government.}\\
Chen Xie (Johns Hopkins University)\\
Jun Hashimoto (Academia Sinica Institute of Astronomy \& Astrophysics)\\
Jingyi Zhang (University of Hawai'i at Manoa)\\
Becca Michelson (Amherst College)\\
Eliot Halley Vrijmoet (Smith College) \\
Christine Chen (Space Telescope Science Institute)\\
Emily Rickman (Space Telescope Science Institute)\\
Kielan Hoch (Space Telescope Science Institute)\\
Kate Follette (Amherst College)
\linebreak

\textbf{Abstract:}
The Space Telescope Imaging Spectrograph (STIS) on the \textit{Hubble Space Telescope} currently stands as the sole space-based astronomical facility providing visible-light coronagraphic imaging -- and the only facility \textit{anywhere} that can perform both visible- and ultraviolet-light coronagraphic \textit{spectroscopy}. In imaging, STIS offers unparalleled stability that rivals the performance of ground-based direct imaging in the optical, and a wide field of view that will complement the upcoming capabilities of \textit{Roman} coronagraphy. STIS also has the capability for direct high-contrast visible and ultraviolet spectroscopy via two occulting bars in its 52''$\times$0.2'' spectroscopic slit. By placing a bright astrophysical source behind an occulting bar, it is possible to use the STIS visible and NUV/FUV gratings to obtain spatially-resolved spectra of faint environments and companions, covering wavelengths from 1150-10,300Å at resolutions of $R\sim500-10{,}000$. In this white paper, we detail the use cases and performance of this under-utilized mode, with starlight subtraction enabling visible light spectral contrasts of $\sim10^{-4}-10^{-5}$. We describe the promise of STIS coronagraphic spectroscopy for a wide variety of astrophysical applications -- planetary and brown dwarf companions, circumstellar disks, young stellar objects, evolved stars and binaries, and active galactic nuclei/galaxy host environments -- throughout the 2030s. STIS high-contrast UV spectroscopy in particular could provide transformative science while pathfinding both techniques and observational studies for the Habitable Worlds Observatory.
}
\pagebreak

\setcounter{page}{1}

\textit{\textbf{Background}:} 
The ability to perform imaging and spectroscopy at high contrast ratios between bright astrophysical sources and their faint surroundings has transformed multiple disciplines of astronomy, allowing the direct detection of exoplanets (\citealt{Marois2008}, see \citealt{Currie2023} for a review), constraining the composition of circumstellar disks \citep{schneider99}, tracing kinematics and outflows of young stellar objects \citep{dougados00}, determining the composition of evolved stars \citep{kastner94}, revealing the surroundings of quasar host galaxies \citep{guyon06}, and more. Adaptive optics (AO) systems (e.g., SPHERE, GPI, SCExAO, MagAO-X) enables these advances on 5-10m ground-based telescopes, primarily at infrared (IR; i.e., 1--5 \SI{}{\um}) wavelengths. Compared to performance in the IR, visible-light image quality is poorer, with typical Strehl ratios (SR) of $\sim$10-40\% at V band vs. 70-90\% at H band \citep{muse,close25}. In both cases, residual stellar halo light decorrelates on minutes to hours timescales, and due to reduced image quality and atmospheric attenuation, visible-light high-contrast \textit{spectroscopy} is even more challenging. However, above the Earth's atmosphere, image quality for space-based facilities like the \textit{Hubble Space Telescope} (HST) can be significantly higher (e.g., SR $\sim$ 70\% to 98\% in the visible to IR), limited only by features like optical aberrations and thermal ``breathing", and highly stable over days to months. \textbf{By performing both high-contrast direct imaging \textit{and} spectroscopy from space at wavelengths shorter than \SI{1}{\um}, HST thus offers capabilities unrivaled by other current and planned facilities.}

\begin{wrapfigure}{R}{0.4\textwidth}
    \centering
    \vspace{-1.45em}
    \includegraphics[width = 0.9\linewidth]{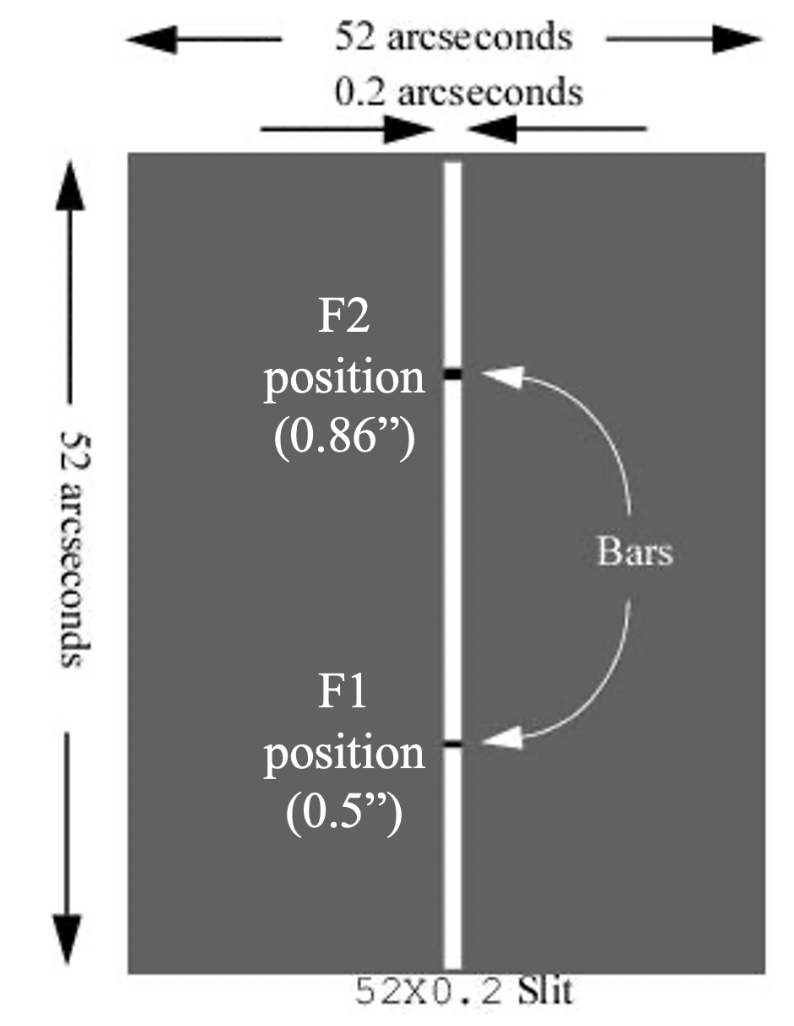}
    \caption{Schematic of the 52$\times$0.2 aperture, illustrating the two fiducial bars used for coronagraphic spectroscopy (image adapted from the STIS Instrument Handbook).}
    \label{fig:bar}
    \vspace{-1em}
\end{wrapfigure}

The Space Telescope Imaging Spectograph (STIS) is HST's visible-light workhorse coronagraphic instrument.  In visible light, STIS imaging coronagraphy achieves contrasts competitive with 8-10m ground-based facilities \citep{debes19}.  Compared to ground-based instruments, HST/STIS also achieves superior raw sensitivity and image quality even for optically faint stars where ground-based AO systems yield degraded performance \citep[typically $V\gtrsim 9-10$; e.g.,][]{macintosh14,beuzit19}. Its highly stable point spread function (PSF) means that the stellar halo can be effectively suppressed even with relatively simple PSF subtraction methods.  

In this white paper, we discuss the STIS coronagraphic capabilities that extend beyond direct imaging to its barred slit spectroscopic modes. Placing a bright object behind one of the barred slits suppresses its light, as shown in Figure~\ref{fig:bar}, allowing high dynamic range, spatially-resolved spectroscopy. \textbf{While this STIS coronagraphic spectroscopy mode has long been available, it has been under-utilized for astrophysics, holding exciting potential for transformative science through the 2030s. } We detail the current understanding of this mode's performance at visible wavelengths, the prospect of extending this method to the ultraviolet (UV), and describe specific science cases that its use can enable (encompassing stellar astrophysics, exoplanets and planet formation, and both galactic and extragalactic astronomy).

\textit{\textbf{Description of STIS modes and previous use}:} STIS is the oldest operational instrument on HST, installed in February 1997 during Servicing Mission 2 (STS-82). STIS features an ensemble of coronagraphic imaging occulters in a focal plane mask (50CORON) that includes bars and wedges, which occult starlight prior to reaching the visible/NUV charge-coupled device (CCD) or the NUV/FUV Multi-Anode Microchannel Array (MAMA) detectors \citep{woodgate98}. In addition to the focal plane masks for direct imaging, occulting bars were also included within one of the STIS spectroscopic slits, the 52'' $\times$ 0.2'' slit. Figure~\ref{fig:bar} illustrates the two occulting bars of widths 0.5'' and 0.86'' located within this slit. The occulting bars can be used with either the CCD or MAMA detectors, and with any of the first-order gratings accessible with STIS long-slit modes --- this provides access to wavelength coverage from 1150--10,300Å at resolutions of $R\sim500-10{,}000$ (i.e., from G140M/L using the MAMAs to G750M/L using the CCD). No other current ground- or space-based instrument has comparable wavelength grasp and resolution with coronagraphic capabilities, motivating the focus of this HST white paper.

Previous science studies using the STIS spectroscopic occulting bars have focused largely on resolved circumstellar disk science, owing to the significant contrast ratio between bright young stars and their faint debris disks (we detail these studies below under ``Key science''). Disk studies \citep{roberge05, Lomax18} have readily achieved \textit{spectral} contrasts (i.e., disk-to-star flux ratios) of $10^{-3}$ to $10^{-4}$ interior to 1'', demonstrating the unique spectroscopic sensitivity achievable with this mode. In total, over 20 HST programs have used the STIS barred spectroscopic modes. These have predominantly used the CCD, although a handful of UV programs have also used the bars with MAMA detectors. However, previous use of coronagraphic spectroscopy has been limited by the postprocessing challenge of PSF subtraction in dispersed 2D spectra. Fortunately, advances in PSF subtraction developed for high-contrast direct imaging can now be applied to coronagraphic spectroscopy, providing a renewed opportunity to use these modes for science. \textbf{In particular, by combining coronagraphic bright star/source suppression with UV sensitivity, further developing these modes in the coming decade will generate exciting, pathfinding UV science in advance of the Habitable Worlds Observatory (HWO).} 

\begin{wrapfigure}{R}{0.45\textwidth}
    \centering
    \vspace{-1.25em}
    \includegraphics[width = 0.9\linewidth]{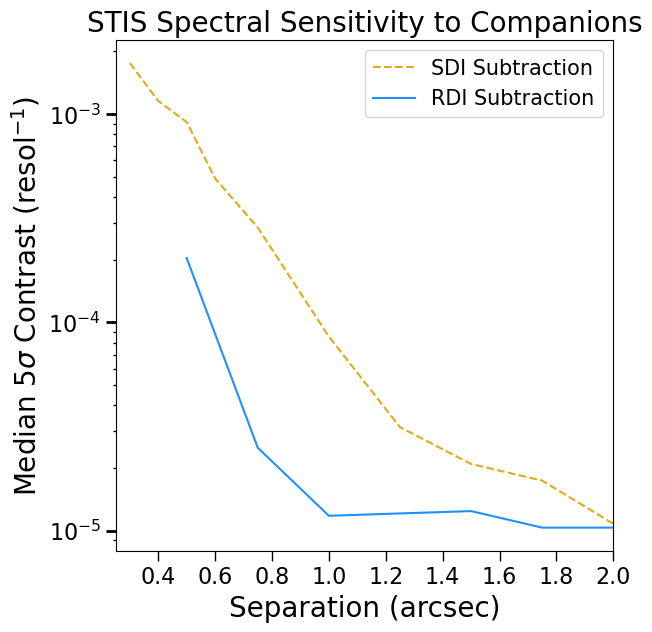}
     \vspace{-1.25em}
    \caption{Spectral contrast performance for both RDI and SDI subtraction methods.}
    \label{fig:contrast}
    \vspace{-0.5em}
\end{wrapfigure}

\textit{\textbf{Current contrast performance:}} Determining achievable spectral contrast limits for coronagraphic spectroscopy requires many areas of optimization (e.g., star centering behind the bar and image referencing, scaling of the reference PSF to match the science target, 2D spectral extraction techniques at close inner working angles dominated by residual speckle noise, etc.). The best contrast performance to-date has been demonstrated in debris disk studies \citep[e.g.,][]{roberge05}, and analyses of archival datasets offer ideal opportunities to characterize the mode and determine best practices (also the focus of recent HST calibration programs, e.g., GO-17092). 

Here, we briefly summarize two approaches to starlight/PSF subtraction and current best performance estimates, which will be described further in a forthcoming work (J. Zhang, \textit{in prep.}). The first approach, classical reference differential imaging \citep[RDI;][]{nicmos_rdi} uses dedicated reference star observations to scale and subtract the PSF, making efforts to match spectral type, brightness, and detector properties as closely as possible between science and reference observations. The second approach, spectral differential imaging \citep[SDI;][]{racine99,marois05,biller06}, leverages the wavelength and spatial diversity of the central PSF and its surroundings without requiring a reference star. By rescaling the image in wavelength space, the central PSF can be effectively masked and interpolated over in the science image to produce a PSF model for clean subtraction, leaving only residual light from the companion/immediate environment. We show preliminary $5\sigma$ contrast curves using the SDI and RDI methods in Figure~\ref{fig:contrast}, based upon observations using the G750L mode (5240--10,270Å) and 0.5'' occulting bar, and calculated as the median contrast across all wavelength channels (e.g., each wavelength sensitive to detecting emission $\sim$10,000$\times$ fainter than the central occulted source at that separation). The two techniques offer different advantages --- RDI providing greater contrast, and SDI providing greater observing efficiency (while mitigating need for a dedicated reference star). RDI and SDI can be combined with powerful PSF subtraction algorithms to further enhance contrast \citep[e.g.,][]{Soummer2012}.

\textbf{\textit{Prospects for UV high-contrast science:}} PSF subtraction techniques have so far been applied to the STIS visible grating modes (G430L/G750L), but both RDI and SDI approaches should also be readily extensible to UV observations with the MAMAs, with potentially improved contrast at closer inner working angles owing to smaller PSF size at shorter wavelengths. If detector health and safety due to host star brightness is a concern, STIS also has the NUV G230LB and G230MB gratings which use the CCD. While CCD NUV sensitivity is moderately lower, it is still sufficient to perform similarly in the visible, enabling NUV spectra for a range of astrophysical objects.\\

\textit{\textbf{Key science enabled with high-contrast optical \& UV spectroscopy}} 

\textbullet\ \textit{Accreting Protoplanets and Substellar Objects:} H$\alpha$ line emission has played an indispensible role in identifying and characterizing accretion properties of protoplanets \citep{Haffert2019,Close2025,Currie2022}.  H$\alpha$ lines due to protoplanet emission may exhibit substantial luminosity variability and exhibit complex shapes like inverse P Cygni profiles diagnostic of the line formation conditions \citep[e.g.,][]{Zhou2025,Currie2025}. STIS coronagraphic spectroscopy can resolve the H$\alpha$ line at twice the resolution achievable with current ground-based instruments (cf. G750M at $R\sim6000$ vs. VLT/MUSE at $R\sim3000$) and with a far more stable PSF.  The NUV can probe continuum emission originating from high-density, compact accretion shocks \citep{Zhou2021}: STIS's stable PSF and deep raw contrast enabled by its coronagraph could, for the first time, probe the continuum in spectroscopy, not narrowband photometry. While there are few direct estimates of accretion rates into the planetary mass regime, the simultaneous detection of H$\alpha$ line emission and NUV continuum from protoplanets will constrain growth, allowing us to assess whether they follow line luminosity vs. accretion rate trends matching those of T Tauri stars, which have been well-explained by multi-component magnetospheric accretion flows \citep[e.g.,][]{Ingleby2013Multiple-flows,Takasao2022MHD3DMA,Ji2026TWHya}.  Additionally, STIS barred spectroscopy enables atmospheric characterization of the brightest directly-imaged young brown dwarf companions at shorter optical wavelengths, covering alkali lines and gravity sensitive molecular features in the red optical \citep[e.g.,][]{mcgovern04}.

\begin{figure}[h!]
    \centering
    \vspace{-0.7em}
    \includegraphics[width = 0.85\linewidth]{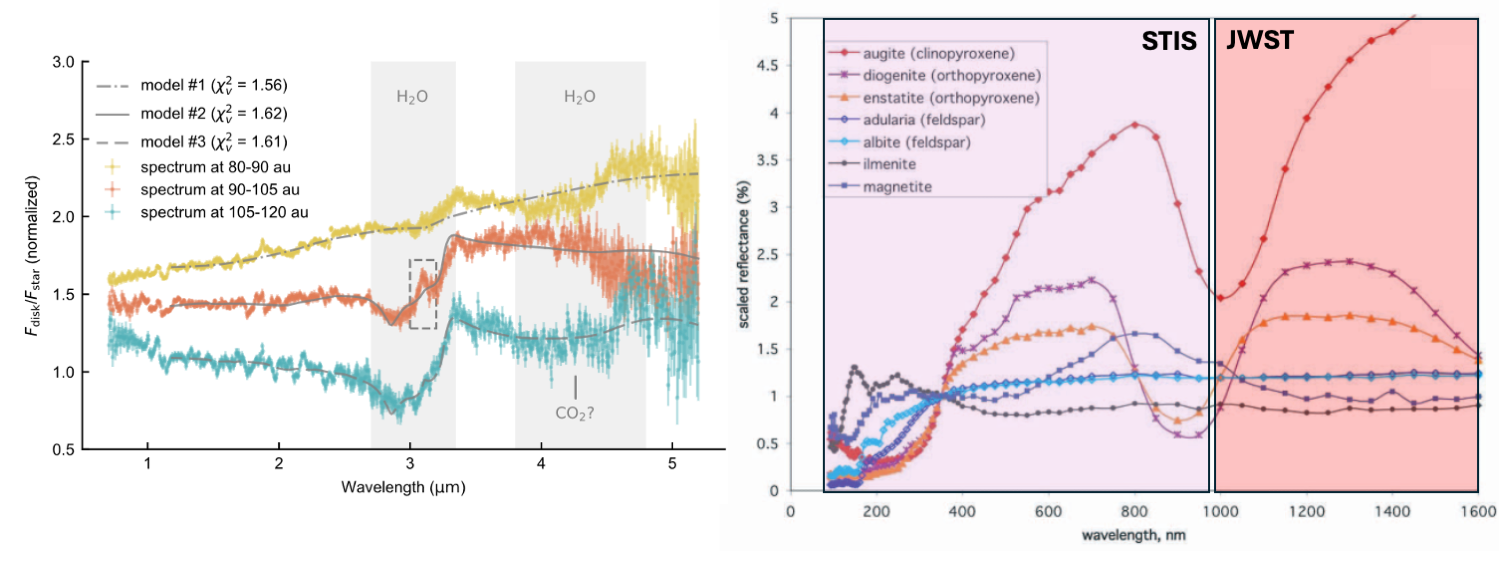}
    \vspace{-1.25em}
    \caption{Combining JWST debris disk spectra \citep[\textit{left};][]{Xie2025}  with NUV disk reflectance spectra 
    \citep[\textit{right};][NUV asteroid spectra]{hendrix06} can constrain dust composition.}
    \label{fig:debris}
    \vspace{-1em}
\end{figure}

\textbullet\ \textit{Protoplanetary and debris disks:}
    Reflectance spectroscopy measures how disk dust modifies stellar light, using wavelength-dependent imprints to constrain dust grain sizes, size distributions, composition, and albedo. STIS coronagraphic spectroscopy also enables searches for UV and visible dust and ice diagnostics known from Solar System objects and laboratory studies, including far-UV ice absorption edges from H$_2$O near 1650~\AA\ and NH$_3$ near $\sim$2000~\AA\ \citep{Hendrix2010, Zastrow2012}, Fe--O charge-transfer absorption at $\sim$2000--4000~\AA\ \citep{Cloutis2008}, and broad Fe$^{2+}$ absorption in silicate minerals, especially orthopyroxene around 0.9~$\mu$m \citep[e.g.,][]{Cloutis2002, Rucks2022}. For protoplanetary disks, observing strategy depends strongly on disk geometry. Low-inclination systems require a fiducial bar (e.g., see STIS spectroscopy of TW~Hya, \citealt{roberge05}). Edge-on disks can naturally occult the star without requiring a bar: in this case, the spectrum can also carry absorption features imprinted along multiple light paths \citep{Hartigan2007, Sturm2023}. For debris disks, coronagraphic spectroscopy probes collisionally produced grains, connecting disk color and spectral features to the dust properties of parent planetesimals \citep[e.g., AU~Mic;][]{Lomax18}. One novel synergy would be to use STIS to observe debris disks with JWST NIR spectra that show ice features, thereby searching for the same types of ices and/or evidence of silicates and iron in NUV reflectance spectra (see Figure \ref{fig:debris}). HST/STIS could obtain the first-ever spatially resolved NUV reflectance spectra of debris disks and protoplanetary disks, opening a new observational window into planet formation and evolution.

\textbullet\ \textit{Young stellar objects:} Young stellar objects (YSOs) are well-suited for high-contrast spectroscopy, which can access jet-launching regions and trace collimated emission or knots along the jet axis \citep[e.g.,][]{Bacciotti00DGTau, Woitas02RWAur, Perrin07LkH233}. Spatially-resolved spectra can yield composition, shock, and kinematic information, via forbidden line diagnostics like [O I] and [S II] \citep{Hartigan95diskaccretion,Bacciotti99HHjets, Flores-Rivera23, Nisini24}, while C IV and Si IV trace hot material in the innermost accretion/ejection zone \citep{Ardila13, Schneider13}, and are particularly well-suited for the innermost regions near bright Herbig Ae/Be stars that would otherwise be inaccessible without coronagraphy. 

\textbullet\ \textit{Stars, binaries, and white dwarfs:} STIS FUV barred spectroscopy has previously been used to investigate interactions between stellar wind and the interstellar medium \citep[ISM;][]{stis_uv_lyalpha}, and high-contrast UV and optical spectroscopy can provide valuable information for both single and multiple star systems. Obtaining spectra of close-angular separation stellar binaries provides compositional and spectral variability information otherwise inaccessible without high contrast, as well as fundamental stellar parameters; one example is determining the mass radius relationship of white dwarfs in Sirius-like systems \citep{zhang23}. Resolved STIS spectra could also provide coarse radial velocity measurements of close companions, effectively making SB1 systems into SB2s.  In terms of evolved stars, red supergiants (RSGs) are likely the dominant source of ISM dust in certain environments; estimates suggest $\approx3 \times 10^{-8}$ M$_{\odot}$ yr$^{-1}$ kpc$^{-2}$ of dust may be injected into the ISM by RSGs in starburst galaxies with large lookback times where carbon-rich Wolf-Rayet (WR) stars are intrinsically rare and AGB stars have not yet formed \citep{Massey03,Levesque2010}. Additionally, dust production rates for some galactic carbon-rich WRs are similar to AGB stars \citep{Lau2020}, but their estimated dust production rates are highly dependent on adopted grain size distributions (e.g., \citet{Wu2026}). Therefore, spatially-resolved spectroscopy of evolved circumstellar environments can determine the rate, velocity, and distribution of RSG and WR mass loss, as well as the resulting circumstellar dust's size and composition, critical to properly account for dust effects when observing the ISM environments of high-redshift galaxies.

\begin{wrapfigure}{R}{0.4\textwidth}
    \centering
    \vspace{-1em}
    \includegraphics[width = 0.9\linewidth]{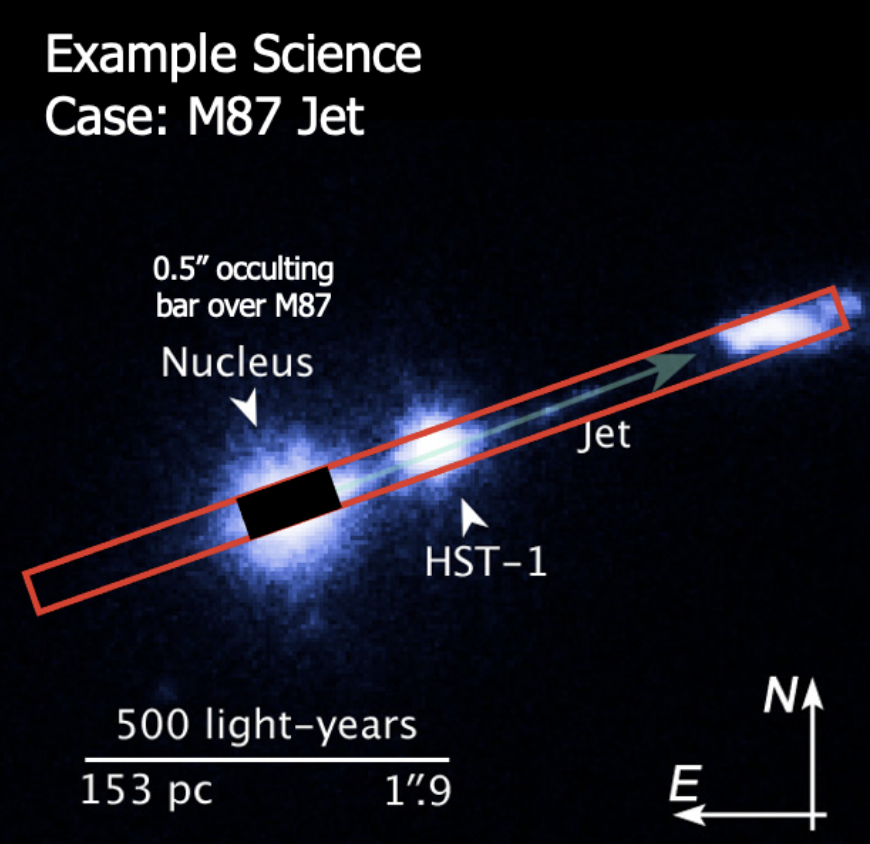}
        \vspace{-1em}
    \caption{The flaring jet from the AGN of M87, illustrating how knots and flares within the jet are amenable to barred spectroscopy. }
    \label{fig:agn}
    \vspace{-0.75em}
\end{wrapfigure}

\textbullet\ \textit{Active galactic nuclei (AGN):} At sub-arcsecond scales, STIS coronagraphic spectroscopy can help separate central AGN emission from galactic stellar populations (e.g., Figure~\ref{fig:agn} adapted from \citealt{madrid09}), as spatially-resolved coronagraphic spectra can suppress the AGN nucleus and trace nearby UV-to-optical continua, and search for emission line ratios that distinguish shock and photoionization \citep{agn}. 

\textit{\textbf{Complementarity with current and future facilities:}} Using HST to extend modern high-contrast techniques to optical and UV wavelengths will play a key complementary role with both ongoing and planned high-contrast observations at other wavelengths and capabilities. For example, the \textbf{Roman Coronagraphic Instrument} (CGI) has a slit spectroscopic mode with bands centered at 660~nm and 730~nm with $R\sim50$ resolution \citep{Kasdin2020}. CGI slit spectroscopy will reach far deeper contrasts ($10^{-8} - 10^{-9}$) than STIS, but only for bright stars (V $\lesssim$ 5--7), at much lower spectral resolution, over narrower bandpasses, and covering smaller angular extents, leaving an important niche for fainter targets to HST. 

At longer wavelengths, the instrumentation of \textbf{JWST} opens up significant opportunities for complementary multi-wavelength studies between the NIR/MIR and UV/optical. There is even limited overlap between the shortest wavelengths accessible to JWST and the longest accessible with STIS: JWST NIRSpec/PRISM and NIRISS/SOSS both have sensitivity down to 0.6$\mu$m, but at significantly lower spectral resolutions ($R\sim30$ and $R\sim700$, respectively) than the STIS medium resolution first-order gratings. However, neither NIRSpec/PRISM nor SOSS have optical elements to suppress starlight, precluding high-contrast observations. 

At the shortest optical and NUV wavelengths, STIS will continue to play an integral role until the advent of \textbf{HWO}. NASA's upcoming Ultraviolet Explorer (\textbf{UVEX}) mission will provide sensitive non-coronagraphic spectroscopy in the NUV and FUV, albeit at much lower spatial resolution than HST. As such, only HST will feature high-contrast UV spectroscopy for the foreseeable future. Efficient, sensitive UV detectors are central to the science planned for HWO, but technology maturation and requirement definitions are still underway. Up until now, what counts for high-contrast UV spectroscopy has consisted mostly of measuring absorption spectra of stellar UV emission through highly-inclined disks \citep[see citations in][]{hwo_uv_tech}. Understanding high-contrast UV performance with STIS (e.g., time-resolved MAMA readouts to characterize speckle noise, observatory thermal and focus impact on PSF stability, determining criteria for reference star observations, etc.) can provide an important bridge across the technology gap between HST and HWO. Coronagraphic spectroscopy at UV wavelengths, in the vein of \citet{stis_uv_lyalpha}, but now with the advantage of a quarter-century of progress in high-contrast post-processing techniques, could be invaluable for establishing the most promising pathways for HWO UV science.

\clearpage

\vspace{5mm}

\section*{References}

\setlength{\bibsep}{-1pt plus 0.25ex}
\renewcommand\refname{\vskip -8mm}
\footnotesize{
\begin{multicols}{2}
{\bibliography{refs.bib}}

@ARTICLE{zhang23,
       author = {{Zhang}, Hengyue and {Brandt}, Timothy D. and {Kiman}, Rocio and {Venner}, Alexander and {An}, Qier and {Chen}, Minghan and {Li}, Yiting},
        title = "{Dynamical masses and ages of Sirius-like systems}",
      journal = {\mnras},
         year = 2023,
        month = sep,
       volume = {524},
       number = {1},
        pages = {695-715},
          doi = {10.1093/mnras/stad1849},
archivePrefix = {arXiv},
       eprint = {2303.08198},
 primaryClass = {astro-ph.SR},
       adsurl = {https://ui.adsabs.harvard.edu/abs/2023MNRAS.524..695Z}
}

@ARTICLE{Close2025,
       author = {{Close}, Laird M. and {van Capelleveen}, Richelle F. and {Weible}, Gabriel and {Wagner}, Kevin and {Haffert}, Sebastiaan Y. and {Males}, Jared R. and {Ilyin}, Ilya and {Kenworthy}, Matthew A. and {Li}, Jialin and {Long}, Joseph D. and {Ertel}, Steve and {Ginski}, Christian and {Weinberger}, Alycia J. and {Follette}, Kate and {Liberman}, Joshua and {Twitchell}, Katie and {Johnson}, Parker and {Kueny}, Jay and {Apai}, Daniel and {Doyon}, Rene and {Foster}, Warren and {Gasho}, Victor and {Van Gorkom}, Kyle and {Guyon}, Olivier and {Kautz}, Maggie Y. and {McLeod}, Avalon and {McEwen}, Eden and {Pearce}, Logan and {Schatz}, Lauren and {Hedglen}, Alexander D. and {Wu}, Ya-Lin and {Isbell}, Jacob and {Power}, Jenny and {Carlson}, Jared and {Close}, Emmeline and {Tonucci}, Elena and {Mars}, Matthijs},
        title = "{Wide Separation Planets in Time (WISPIT): Discovery of a Gap H{\ensuremath{\alpha}} Protoplanet WISPIT 2b with MagAO-X}",
      journal = {\apjl},
         year = 2025,
        month = sep,
       volume = {990},
       number = {1},
          eid = {L9},
        pages = {L9},
          doi = {10.3847/2041-8213/adf7a5},
archivePrefix = {arXiv},
       eprint = {2508.19046},
 primaryClass = {astro-ph.EP},
       adsurl = {https://ui.adsabs.harvard.edu/abs/2025ApJ...990L...9C}
}

@ARTICLE{agn,
       author = {{Kraemer}, Steven B. and {Crenshaw}, D. Michael},
        title = "{Resolved Spectroscopy of the Narrow-Line Region in NGC 1068. III. Physical Conditions in the Emission-Line Gas}",
      journal = {\apj},
         year = 2000,
        month = dec,
       volume = {544},
       number = {2},
        pages = {763-779},
          doi = {10.1086/317246},
archivePrefix = {arXiv},
       eprint = {astro-ph/0007018},
 primaryClass = {astro-ph},
       adsurl = {https://ui.adsabs.harvard.edu/abs/2000ApJ...544..763K}
}

@ARTICLE{Currie2022,
       author = {{Currie}, Thayne and {Lawson}, Kellen and {Schneider}, Glenn and {Lyra}, Wladimir and {Wisniewski}, John and {Grady}, Carol and {Guyon}, Olivier and {Tamura}, Motohide and {Kotani}, Takayuki and {Kawahara}, Hajime and {Brandt}, Timothy and {Uyama}, Taichi and {Muto}, Takayuki and {Dong}, Ruobing and {Kudo}, Tomoyuki and {Hashimoto}, Jun and {Fukagawa}, Misato and {Wagner}, Kevin and {Lozi}, Julien and {Chilcote}, Jeffrey and {Tobin}, Taylor and {Groff}, Tyler and {Ward-Duong}, Kimberly and {Januszewski}, William and {Norris}, Barnaby and {Tuthill}, Peter and {van der Marel}, Nienke and {Sitko}, Michael and {Deo}, Vincent and {Vievard}, Sebastien and {Jovanovic}, Nemanja and {Martinache}, Frantz and {Skaf}, Nour},
        title = "{Images of embedded Jovian planet formation at a wide separation around AB Aurigae}",
      journal = {Nature Astronomy},
         year = 2022,
        month = apr,
       volume = {6},
        pages = {751-759},
          doi = {10.1038/s41550-022-01634-x},
archivePrefix = {arXiv},
       eprint = {2204.00633},
 primaryClass = {astro-ph.EP},
       adsurl = {https://ui.adsabs.harvard.edu/abs/2022NatAs...6..751C}
}

@ARTICLE{Currie2025,
       author = {{Currie}, Thayne and {Hashimoto}, Jun and {Aoyama}, Yuhiko and {Dong}, Ruobing and {Fukagawa}, Misato and {Muto}, Takayuki and {Dykes}, Erica and {El Morsy}, Mona and {Tamura}, Motohide},
        title = "{VLT/MUSE Detection of the AB Aurigae b Protoplanet with H$_{{\ensuremath{\alpha}}}$ Spectroscopy}",
      journal = {\apjl},
         year = 2025,
        month = sep,
       volume = {990},
       number = {2},
          eid = {L42},
        pages = {L42},
          doi = {10.3847/2041-8213/adf7a0},
archivePrefix = {arXiv},
       eprint = {2508.18351},
 primaryClass = {astro-ph.EP},
       adsurl = {https://ui.adsabs.harvard.edu/abs/2025ApJ...990L..42C}
}

@ARTICLE{mcgovern04,
       author = {{McGovern}, Mark R. and {Kirkpatrick}, J. Davy and {McLean}, Ian S. and {Burgasser}, Adam J. and {Prato}, L. and {Lowrance}, Patrick J.},
        title = "{Identifying Young Brown Dwarfs Using Gravity-Sensitive Spectral Features}",
      journal = {\apj},
         year = 2004,
        month = jan,
       volume = {600},
       number = {2},
        pages = {1020-1024},
          doi = {10.1086/379849},
archivePrefix = {arXiv},
       eprint = {astro-ph/0309634},
 primaryClass = {astro-ph},
       adsurl = {https://ui.adsabs.harvard.edu/abs/2004ApJ...600.1020M}
}

@INPROCEEDINGS{Currie2023,
       author = {{Currie}, T. and {Biller}, B. and {Lagrange}, A. and {Marois}, C. and {Guyon}, O. and {Nielsen}, E.~L. and {Bonnefoy}, M. and {De Rosa}, R.~J.},
        title = "{Direct Imaging and Spectroscopy of Extrasolar Planets}",
    booktitle = {Protostars and Planets VII},
         year = 2023,
       editor = {{Inutsuka}, S. and {Aikawa}, Y. and {Muto}, T. and {Tomida}, K. and {Tamura}, M.},
       series = {Astronomical Society of the Pacific Conference Series},
       volume = {534},
        month = jul,
        pages = {799},
          doi = {10.48550/arXiv.2205.05696},
archivePrefix = {arXiv},
       eprint = {2205.05696},
 primaryClass = {astro-ph.EP},
       adsurl = {https://ui.adsabs.harvard.edu/abs/2023ASPC..534..799C}
}

@ARTICLE{Haffert2019,
       author = {{Haffert}, S.~Y. and {Bohn}, A.~J. and {de Boer}, J. and {Snellen}, I.~A.~G. and {Brinchmann}, J. and {Girard}, J.~H. and {Keller}, C.~U. and {Bacon}, R.},
        title = "{Two accreting protoplanets around the young star PDS 70}",
      journal = {Nature Astronomy},
         year = 2019,
        month = jun,
       volume = {3},
        pages = {749-754},
          doi = {10.1038/s41550-019-0780-5},
archivePrefix = {arXiv},
       eprint = {1906.01486},
 primaryClass = {astro-ph.EP},
       adsurl = {https://ui.adsabs.harvard.edu/abs/2019NatAs...3..749H}
}

@ARTICLE{Marois2008,
       author = {{Marois}, Christian and {Macintosh}, Bruce and {Barman}, Travis and {Zuckerman}, B. and {Song}, Inseok and {Patience}, Jennifer and {Lafreni{\`e}re}, David and {Doyon}, Ren{\'e}},
        title = "{Direct Imaging of Multiple Planets Orbiting the Star HR 8799}",
      journal = {Science},
         year = 2008,
        month = nov,
       volume = {322},
       number = {5906},
        pages = {1348},
          doi = {10.1126/science.1166585},
archivePrefix = {arXiv},
       eprint = {0811.2606},
 primaryClass = {astro-ph},
       adsurl = {https://ui.adsabs.harvard.edu/abs/2008Sci...322.1348M}
}

@ARTICLE{Soummer2012,
       author = {{Soummer}, R{\'e}mi and {Pueyo}, Laurent and {Larkin}, James},
        title = "{Detection and Characterization of Exoplanets and Disks Using Projections on Karhunen-Lo{\`e}ve Eigenimages}",
      journal = {\apjl},
         year = 2012,
        month = aug,
       volume = {755},
       number = {2},
          eid = {L28},
        pages = {L28},
          doi = {10.1088/2041-8205/755/2/L28},
archivePrefix = {arXiv},
       eprint = {1207.4197},
 primaryClass = {astro-ph.IM},
       adsurl = {https://ui.adsabs.harvard.edu/abs/2012ApJ...755L..28S}
}

@ARTICLE{Zhou2021,
       author = {{Zhou}, Yifan and {Bowler}, Brendan P. and {Wagner}, Kevin R. and {Schneider}, Glenn and {Apai}, D{\'a}niel and {Kraus}, Adam L. and {Close}, Laird M. and {Herczeg}, Gregory J. and {Fang}, Min},
        title = "{Hubble Space Telescope UV and H{\ensuremath{\alpha}} Measurements of the Accretion Excess Emission from the Young Giant Planet PDS 70 b}",
      journal = {\aj},
         year = 2021,
        month = may,
       volume = {161},
       number = {5},
          eid = {244},
        pages = {244},
          doi = {10.3847/1538-3881/abeb7a},
archivePrefix = {arXiv},
       eprint = {2104.13934},
 primaryClass = {astro-ph.EP},
       adsurl = {https://ui.adsabs.harvard.edu/abs/2021AJ....161..244Z}
}

@ARTICLE{Zhou2025,
       author = {{Zhou}, Yifan and {Bowler}, Brendan P. and {Sanghi}, Aniket and {Marleau}, Gabriel-Dominique and {Takasao}, Shinsuke and {Aoyama}, Yuhiko and {Hasegawa}, Yasuhiro and {Thanathibodee}, Thanawuth and {Uyama}, Taichi and {Hashimoto}, Jun and {Wagner}, Kevin and {Calvet}, Nuria and {Demars}, Dorian and {Wu}, Ya-Lin and {Biddle}, Lauren I. and {Haffert}, Sebastiaan Y. and {Bryan}, Marta L.},
        title = "{Evidence for Variable Accretion onto PDS 70 c and Implications for Protoplanet Detections}",
      journal = {\apjl},
         year = 2025,
        month = feb,
       volume = {980},
       number = {2},
          eid = {L39},
        pages = {L39},
          doi = {10.3847/2041-8213/adb134},
archivePrefix = {arXiv},
       eprint = {2502.14024},
 primaryClass = {astro-ph.EP},
       adsurl = {https://ui.adsabs.harvard.edu/abs/2025ApJ...980L..39Z}
}

@ARTICLE{hendrix06,
       author = {{Hendrix}, Amanda R. and {Vilas}, Faith},
        title = "{The Effects of Space Weathering at UV Wavelengths: S-Class Asteroids}",
      journal = {\aj},
         year = 2006,
        month = sep,
       volume = {132},
       number = {3},
        pages = {1396-1404},
          doi = {10.1086/506426},
       adsurl = {https://ui.adsabs.harvard.edu/abs/2006AJ....132.1396H}
}

@INPROCEEDINGS{Kasdin2020,
       author = {{Kasdin}, N. Jeremy and {Bailey}, Vanessa P. and {Mennesson}, Bertrand and {Zellem}, Robert T. and {Ygouf}, Marie and {Rhodes}, Jason and {Luchik}, Thomas and {Zhao}, Feng and {Riggs}, A.~J. Eldorado and {Seo}, Byoung-Joon and {Krist}, John and {Kern}, Brian and {Tang}, Hong and {Nemati}, Bijan and {Groff}, Tyler D. and {Zimmerman}, Neil and {Macintosh}, Bruce and {Turnbull}, Margaret and {Debes}, John and {Douglas}, Ewan S. and {Lupu}, Roxana E.},
        title = "{The Nancy Grace Roman Space Telescope Coronagraph Instrument (CGI) technology demonstration}",
    booktitle = {Space Telescopes and Instrumentation 2020: Optical, Infrared, and Millimeter Wave},
         year = 2020,
       editor = {{Lystrup}, Makenzie and {Perrin}, Marshall D.},
       series = {Society of Photo-Optical Instrumentation Engineers (SPIE) Conference Series},
       volume = {11443},
        month = dec,
          eid = {114431U},
        pages = {114431U},
          doi = {10.1117/12.2562997},
archivePrefix = {arXiv},
       eprint = {2103.01980},
 primaryClass = {astro-ph.IM},
       adsurl = {https://ui.adsabs.harvard.edu/abs/2020SPIE11443E..1UK}
}

@ARTICLE{Levesque2010,
       author = {{Levesque}, Emily M.},
        title = "{The physical properties of red supergiants}",
      journal = {\nar},
         year = 2010,
        month = jan,
       volume = {54},
       number = {1-2},
        pages = {1-12},
          doi = {10.1016/j.newar.2009.10.002},
archivePrefix = {arXiv},
       eprint = {0902.2789},
 primaryClass = {astro-ph.SR},
       adsurl = {https://ui.adsabs.harvard.edu/abs/2010NewAR..54....1L}
}

@ARTICLE{madrid09,
       author = {{Madrid}, Juan P.},
        title = "{Hubble Space Telescope Observations of an Extraordinary Flare in the M87 Jet}",
      journal = {\aj},
         year = 2009,
        month = apr,
       volume = {137},
       number = {4},
        pages = {3864-3868},
          doi = {10.1088/0004-6256/137/4/3864},
archivePrefix = {arXiv},
       eprint = {0904.3546},
 primaryClass = {astro-ph.CO},
       adsurl = {https://ui.adsabs.harvard.edu/abs/2009AJ....137.3864M}
}

@ARTICLE{woodgate98,
       author = {{Woodgate}, B.~E. and {Kimble}, R.~A. and {Bowers}, C.~W. and {Kraemer}, S. and {Kaiser}, M.~E. and {Danks}, A.~C. and {Grady}, J.~F. and {Loiacono}, J.~J. and {Brumfield}, M. and {Feinberg}, L. and {Gull}, T.~R. and {Heap}, S.~R. and {Maran}, S.~P. and {Lindler}, D. and {Hood}, D. and {Meyer}, W. and {Vanhouten}, C. and {Argabright}, V. and {Franka}, S. and {Bybee}, R. and {Dorn}, D. and {Bottema}, M. and {Woodruff}, R. and {Michika}, D. and {Sullivan}, J. and {Hetlinger}, J. and {Ludtke}, C. and {Stocker}, R. and {Delamere}, A. and {Rose}, D. and {Becker}, I. and {Garner}, H. and {Timothy}, J.~G. and {Blouke}, M. and {Joseph}, C.~L. and {Hartig}, G. and {Green}, R.~F. and {Jenkins}, E.~B. and {Linsky}, J.~L. and {Hutchings}, J.~B. and {Moos}, H.~W. and {Boggess}, A. and {Roesler}, F. and {Weistrop}, D.},
        title = "{The Space Telescope Imaging Spectrograph Design}",
      journal = {\pasp},
         year = 1998,
        month = oct,
       volume = {110},
       number = {752},
        pages = {1183-1204},
          doi = {10.1086/316243},
       adsurl = {https://ui.adsabs.harvard.edu/abs/1998PASP..110.1183W}
}

@ARTICLE{macintosh14,
       author = {{Macintosh}, Bruce and {Graham}, James R. and {Ingraham}, Patrick and {Konopacky}, Quinn and {Marois}, Christian and {Perrin}, Marshall and {Poyneer}, Lisa and {Bauman}, Brian and {Barman}, Travis and {Burrows}, Adam S. and {Cardwell}, Andrew and {Chilcote}, Jeffrey and {De Rosa}, Robert J. and {Dillon}, Daren and {Doyon}, Rene and {Dunn}, Jennifer and {Erikson}, Darren and {Fitzgerald}, Michael P. and {Gavel}, Donald and {Goodsell}, Stephen and {Hartung}, Markus and {Hibon}, Pascale and {Kalas}, Paul and {Larkin}, James and {Maire}, Jerome and {Marchis}, Franck and {Marley}, Mark S. and {McBride}, James and {Millar-Blanchaer}, Max and {Morzinski}, Katie and {Norton}, Andrew and {Oppenheimer}, B.~R. and {Palmer}, David and {Patience}, Jennifer and {Pueyo}, Laurent and {Rantakyro}, Fredrik and {Sadakuni}, Naru and {Saddlemyer}, Leslie and {Savransky}, Dmitry and {Serio}, Andrew and {Soummer}, Remi and {Sivaramakrishnan}, Anand and {Song}, Inseok and {Thomas}, Sandrine and {Wallace}, J. Kent and {Wiktorowicz}, Sloane and {Wolff}, Schuyler},
        title = "{First light of the Gemini Planet Imager}",
      journal = {Proceedings of the National Academy of Science},
         year = 2014,
        month = sep,
       volume = {111},
       number = {35},
        pages = {12661-12666},
          doi = {10.1073/pnas.1304215111},
archivePrefix = {arXiv},
       eprint = {1403.7520},
 primaryClass = {astro-ph.EP},
       adsurl = {https://ui.adsabs.harvard.edu/abs/2014PNAS..11112661M}
}

@ARTICLE{beuzit19,
       author = {{Beuzit}, J.-L. and {Vigan}, A. and {Mouillet}, D. and {Dohlen}, K. and {Gratton}, R. and {Boccaletti}, A. and {Sauvage}, J.-F. and {Schmid}, H.~M. and {Langlois}, M. and {Petit}, C. and {Baruffolo}, A. and {Feldt}, M. and {Milli}, J. and {Wahhaj}, Z. and {Abe}, L. and {Anselmi}, U. and {Antichi}, J. and {Barette}, R. and {Baudrand}, J. and {Baudoz}, P. and {Bazzon}, A. and {Bernardi}, P. and {Blanchard}, P. and {Brast}, R. and {Bruno}, P. and {Buey}, T. and {Carbillet}, M. and {Carle}, M. and {Cascone}, E. and {Chapron}, F. and {Charton}, J. and {Chauvin}, G. and {Claudi}, R. and {Costille}, A. and {De Caprio}, V. and {de Boer}, J. and {Delboulb{\'e}}, A. and {Desidera}, S. and {Dominik}, C. and {Downing}, M. and {Dupuis}, O. and {Fabron}, C. and {Fantinel}, D. and {Farisato}, G. and {Feautrier}, P. and {Fedrigo}, E. and {Fusco}, T. and {Gigan}, P. and {Ginski}, C. and {Girard}, J. and {Giro}, E. and {Gisler}, D. and {Gluck}, L. and {Gry}, C. and {Henning}, T. and {Hubin}, N. and {Hugot}, E. and {Incorvaia}, S. and {Jaquet}, M. and {Kasper}, M. and {Lagadec}, E. and {Lagrange}, A.-M. and {Le Coroller}, H. and {Le Mignant}, D. and {Le Ruyet}, B. and {Lessio}, G. and {Lizon}, J.-L. and {Llored}, M. and {Lundin}, L. and {Madec}, F. and {Magnard}, Y. and {Marteaud}, M. and {Martinez}, P. and {Maurel}, D. and {M{\'e}nard}, F. and {Mesa}, D. and {M{\"o}ller-Nilsson}, O. and {Moulin}, T. and {Moutou}, C. and {Orign{\'e}}, A. and {Parisot}, J. and {Pavlov}, A. and {Perret}, D. and {Pragt}, J. and {Puget}, P. and {Rabou}, P. and {Ramos}, J. and {Reess}, J.-M. and {Rigal}, F. and {Rochat}, S. and {Roelfsema}, R. and {Rousset}, G. and {Roux}, A. and {Saisse}, M. and {Salasnich}, B. and {Santambrogio}, E. and {Scuderi}, S. and {Segransan}, D. and {Sevin}, A. and {Siebenmorgen}, R. and {Soenke}, C. and {Stadler}, E. and {Suarez}, M. and {Tiph{\`e}ne}, D. and {Turatto}, M. and {Udry}, S. and {Vakili}, F. and {Waters}, L.~B.~F.~M. and {Weber}, L. and {Wildi}, F. and {Zins}, G. and {Zurlo}, A.},
        title = "{SPHERE: the exoplanet imager for the Very Large Telescope}",
      journal = {\aap},
         year = 2019,
        month = nov,
       volume = {631},
          eid = {A155},
        pages = {A155},
          doi = {10.1051/0004-6361/201935251},
archivePrefix = {arXiv},
       eprint = {1902.04080},
 primaryClass = {astro-ph.IM},
       adsurl = {https://ui.adsabs.harvard.edu/abs/2019A&A...631A.155B}
}

@INPROCEEDINGS{muse,
       author = {{Wevers}, T. and {Selman}, F. and {Reyes}, A. and {Vega}, M. and {Hartke}, J. and {Bian}, F. and {Beltramo-Martin}, O. and {F{\'e}tick}, R.~J.~L. and {Kamann}, S. and {Kolb}, J. and {Kravtsov}, T. and {Moya}, C. and {Neichel}, B. and {Oberti}, S. and {Reyes}, C. and {Valenti}, E.},
        title = "{Performance characterization and near-real-time monitoring of MUSE adaptive optics modes at Paranal}",
    booktitle = {Observatory Operations: Strategies, Processes, and Systems IX},
         year = 2022,
       editor = {{Adler}, David S. and {Seaman}, Robert L. and {Benn}, Chris R.},
       series = {Society of Photo-Optical Instrumentation Engineers (SPIE) Conference Series},
       volume = {12186},
        month = aug,
          eid = {121860T},
        pages = {121860T},
          doi = {10.1117/12.2630835},
archivePrefix = {arXiv},
       eprint = {2209.07540},
 primaryClass = {astro-ph.IM},
       adsurl = {https://ui.adsabs.harvard.edu/abs/2022SPIE12186E..0TW}
}

@ARTICLE{close25,
       author = {{Close}, Laird M. and {van Capelleveen}, Richelle F. and {Weible}, Gabriel and {Wagner}, Kevin and {Haffert}, Sebastiaan Y. and {Males}, Jared R. and {Ilyin}, Ilya and {Kenworthy}, Matthew A. and {Li}, Jialin and {Long}, Joseph D. and {Ertel}, Steve and {Ginski}, Christian and {Weinberger}, Alycia J. and {Follette}, Kate and {Liberman}, Joshua and {Twitchell}, Katie and {Johnson}, Parker and {Kueny}, Jay and {Apai}, Daniel and {Doyon}, Rene and {Foster}, Warren and {Gasho}, Victor and {Van Gorkom}, Kyle and {Guyon}, Olivier and {Kautz}, Maggie Y. and {McLeod}, Avalon and {McEwen}, Eden and {Pearce}, Logan and {Schatz}, Lauren and {Hedglen}, Alexander D. and {Wu}, Ya-Lin and {Isbell}, Jacob and {Power}, Jenny and {Carlson}, Jared and {Close}, Emmeline and {Tonucci}, Elena and {Mars}, Matthijs},
        title = "{Wide Separation Planets in Time (WISPIT): Discovery of a Gap H{\ensuremath{\alpha}} Protoplanet WISPIT 2b with MagAO-X}",
      journal = {\apjl},
         year = 2025,
        month = sep,
       volume = {990},
       number = {1},
          eid = {L9},
        pages = {L9},
          doi = {10.3847/2041-8213/adf7a5},
archivePrefix = {arXiv},
       eprint = {2508.19046},
 primaryClass = {astro-ph.EP},
       adsurl = {https://ui.adsabs.harvard.edu/abs/2025ApJ...990L...9C}
}

@ARTICLE{Lomax18,
       author = {{Lomax}, Jamie R. and {Wisniewski}, John P. and {Roberge}, Aki and {Donaldson}, Jessica K. and {Debes}, John H. and {Malumuth}, Eliot M. and {Weinberger}, Alycia J.},
        title = "{Optical Coronagraphic Spectroscopy of AU Mic: Evidence of Time Variable Colors?}",
      journal = {\aj},
         year = 2018,
        month = feb,
       volume = {155},
       number = {2},
          eid = {62},
        pages = {62},
          doi = {10.3847/1538-3881/aaa1a7},
archivePrefix = {arXiv},
       eprint = {1705.09291},
 primaryClass = {astro-ph.SR},
       adsurl = {https://ui.adsabs.harvard.edu/abs/2018AJ....155...62L}
}

@ARTICLE{roberge05,
       author = {{Roberge}, Aki and {Weinberger}, Alycia J. and {Malumuth}, Eliot M.},
        title = "{Spatially Resolved Spectroscopy and Coronagraphic Imaging of the TW Hydrae Circumstellar Disk}",
      journal = {\apj},
         year = 2005,
        month = apr,
       volume = {622},
       number = {2},
        pages = {1171-1181},
          doi = {10.1086/427974},
archivePrefix = {arXiv},
       eprint = {astro-ph/0410251},
 primaryClass = {astro-ph},
       adsurl = {https://ui.adsabs.harvard.edu/abs/2005ApJ...622.1171R}
}

@ARTICLE{dougados00,
       author = {{Dougados}, C. and {Cabrit}, S. and {Lavalley}, C. and {M{\'e}nard}, F.},
        title = "{T Tauri stars microjets resolved by adaptive optics}",
      journal = {\aap},
         year = 2000,
        month = may,
       volume = {357},
        pages = {L61-L64},
       adsurl = {https://ui.adsabs.harvard.edu/abs/2000A&A...357L..61D}
}

@ARTICLE{guyon06,
       author = {{Guyon}, O. and {Sanders}, D.~B. and {Stockton}, Alan},
        title = "{Near-Infrared Adaptive Optics Imaging of QSO Host Galaxies}",
      journal = {\apjs},
         year = 2006,
        month = sep,
       volume = {166},
       number = {1},
        pages = {89-127},
          doi = {10.1086/505030},
archivePrefix = {arXiv},
       eprint = {astro-ph/0605079},
 primaryClass = {astro-ph},
       adsurl = {https://ui.adsabs.harvard.edu/abs/2006ApJS..166...89G}
}

@ARTICLE{kastner94,
       author = {{Kastner}, Joel H. and {Weintraub}, David A.},
        title = "{Broken Symmetry: The Structure of the Dust Envelope of IRC +10216}",
      journal = {\apj},
         year = 1994,
        month = oct,
       volume = {434},
        pages = {719},
          doi = {10.1086/174774},
       adsurl = {https://ui.adsabs.harvard.edu/abs/1994ApJ...434..719K}
}

@ARTICLE{schneider99,
       author = {{Schneider}, Glenn and {Smith}, Bradford A. and {Becklin}, E.~E. and {Koerner}, David W. and {Meier}, Roland and {Hines}, Dean C. and {Lowrance}, Patrick J. and {Terrile}, Richard J. and {Thompson}, Rodger I. and {Rieke}, Marcia},
        title = "{NICMOS Imaging of the HR 4796A Circumstellar Disk}",
      journal = {\apjl},
         year = 1999,
        month = mar,
       volume = {513},
       number = {2},
        pages = {L127-L130},
          doi = {10.1086/311921},
archivePrefix = {arXiv},
       eprint = {astro-ph/9901218},
 primaryClass = {astro-ph},
       adsurl = {https://ui.adsabs.harvard.edu/abs/1999ApJ...513L.127S}
}

@ARTICLE{Massey03,
       author = {{Massey}, Philip and {Olsen}, K.~A.~G.},
        title = "{The Evolution of Massive Stars. I. Red Supergiants in the Magellanic Clouds}",
      journal = {\aj},
         year = 2003,
        month = dec,
       volume = {126},
       number = {6},
        pages = {2867-2886},
          doi = {10.1086/379558},
archivePrefix = {arXiv},
       eprint = {astro-ph/0309272},
 primaryClass = {astro-ph},
       adsurl = {https://ui.adsabs.harvard.edu/abs/2003AJ....126.2867M}
}
\bibliographystyle{apj}
\end{multicols}}

\end{document}